\documentclass[%
 reprint,
 superscriptaddress,
 nobibnotes,
 amsmath,amssymb,
 aps,
]{revtex4-1}

\usepackage{graphicx}
\usepackage{dcolumn}
\usepackage{bm}
\usepackage[T1]{fontenc} 
\usepackage[utf8]{inputenc}
\usepackage{subfigure}
\usepackage{hyperref}
\usepackage{xcolor}
\usepackage[english]{babel}

\usepackage{xspace}
\newcommand{\dd}{\ensuremath{\mathrm{d}}\xspace}
\newcommand{\tio}{TiO$_2$\xspace}
\newcommand{\ls}{\ensuremath{\ell^\star}\xspace}
\newcommand{\kls}{\ensuremath{k\ell^\star}\xspace}
\newcommand{\neff}{\ensuremath{n_\mathrm{eff}}\xspace}
\newcommand{\no}{\ensuremath{n_0}\xspace}
\newcommand{\lo}{\ensuremath{\lambda_0}\xspace}
\newcommand{\vE}{\ensuremath{v_\mathrm{E}}\xspace}

\AtEndDocument{%
    \newwrite\bibnotes
    \def\bibnotesext{Notes.bib}
    \immediate\openout\bibnotes=\jobname\bibnotesext
    \immediate\write\bibnotes{@CONTROL{REVTEX41Control}}
    \immediate\write\bibnotes{@CONTROL{%
    apsrev41Control,author="08",editor="1",pages="1",title="0",year="1"}}
     \if@filesw
     \immediate\write\@auxout{\string\citation{apsrev41Control}}%
    \fi
}%

\usepackage{titletoc}

\begin{document}
\title{Tunable high-index photonic glasses}

\author{Lukas Schertel}
\affiliation
{Fachbereich Physik, Universität Konstanz, Universitätsstraße 10, D-78457 Konstanz, Germany}
\affiliation
{Physik-Insitut, Universität Zürich, Winterthurerstrasse 190, CH-8057 Zürich, Switzerland}

\author{Ilona Wimmer}
\affiliation
{Fachbereich Physik, Universität Konstanz, Universitätsstraße 10, D-78457 Konstanz, Germany}
\affiliation
{Fachbereich Chemie, Universität Konstanz, Universitätsstraße 10, D-78457 Konstanz, Germany}

\author{Patricia Besirske}
\affiliation
{Fachbereich Chemie, Universität Konstanz, Universitätsstraße 10, D-78457 Konstanz, Germany}

\author{Christof M. Aegerter}
\affiliation
{Physik-Insitut, Universität Zürich, Winterthurerstrasse 190, CH-8057 Zürich, Switzerland}

\author{Georg Maret}
\affiliation
{Fachbereich Physik, Universität Konstanz, Universitätsstraße 10, D-78457 Konstanz, Germany}

\author{Sebastian Polarz}
\email{sebastian.polarz@uni-konstanz.de}
\affiliation
{Fachbereich Chemie, Universität Konstanz, Universitätsstraße 10, D-78457 Konstanz, Germany}

\author{Geoffroy J. Aubry}
\email{geoffroy.aubry@uni-konstanz.de}
\altaffiliation[Now at ]
{Département de Physique, Université de Fribourg, Chemin du Musée 3, CH-1700 Fribourg, Switzerland}
\affiliation
{Fachbereich Physik, Universität Konstanz, Universitätsstraße 10, D-78457 Konstanz, Germany}

\date{\today}


\begin{abstract}
Materials with extreme photonic properties such as maximum diffuse reflectance, high albedo, or tunable band gaps are essential in many current and future photonic devices and coatings. While photonic crystals, periodic anisotropic structures, are well established, their disordered counterparts, photonic glasses (PGs), are less understood despite their most interesting isotropic photonic properties. Here, we introduce a controlled high index model PG system. It is made of monodisperse spherical TiO$_2$ colloids to exploit strongly resonant Mie scattering for optimal turbidity. We report spectrally resolved combined measurements of turbidity and light energy velocity from large monolithic crack-free samples.  This material class reveals pronounced resonances enabled by the possibility to tune both the refractive index of the extremely low polydisperse constituents and their radius. All our results are rationalized by a model based on the energy coherent potential approximation, which is free of any fitting parameter. Surprisingly good quantitative agreement is found even at high index and elevated packing fraction. This class of PGs may be the key to optimized tunable photonic materials and also central to understand fundamental questions such as isotropic structural colors, random lasing or strong light localization in 3D.
\end{abstract}

\maketitle

\section{Introduction}

The interaction of light with matter is of paramount importance for numerous technologies.
While many of them rely on electronic excitation (e.g., photovoltaics, photocatalysis), it is also highly relevant to control the propagation of light using materials structured at the scale of the optical wavelength.
So far, research on photonic structures mainly addressed photonic crystals (PCs) which are periodic patterns of the refractive index in space~\cite{Yablonovitch1987,*John1987,*Yablonovitch1989,Joannopoulos1997,Lopez2003}.
The lattice structure of PCs implies strongly anisotropic photonic properties such as wavelength-dependent Bragg-scattering, angular-dependent structural coloration in biology~\cite{Vukusic1999,*Vukusic2003}, and anisotropic band gaps.
A well-known strategy for the generation of PCs involves the self-assembly of monodisperse colloidal particles which naturally tend to crystallize~\cite{Freymann2013}. 

Less attention has been paid to photonic glasses (PGs), the disordered counterpart of PCs~\cite{Ballato2000,Garcia2007,*Garcia2010}.
PGs are important because, in many applications such as white paints, coatings, diffusors, or matrix materials for future photonic devices, a prime requirement is the isotropy of photonic band gaps, transport, reflectance, and transmittance. The strongest isotropic scattering materials nowadays, commercial white paints, are a class of empirically optimized turbid materials.
Moreover, PGs can be used for the exploration of various optical phenomena such as Anderson localization of light~\cite{Anderson1985}, random lasing~\cite{Gaio2015}, and tunable isotropic structural colors~\cite{Forster2010,*Park2014,*Xiao2017,*Shang2018}.

Thus, it is of prime interest to understand light transport in high index disordered photonic materials quantitatively by using an appropriate transport theory which can be bench-marked by experiments.
Such analysis relies on model materials which have to fulfill a large number of requirements at once.
The size of the colloidal particles should be of the order of the optical wavelength ($400-800$\,nm).
To fully exploit the enhancement of scattering efficiency due to Mie resonances the size-distribution used for the construction of the materials should be as narrow as possible (polydispersity $< 5\%$) as, otherwise, polydispersity in size and shape smears out resonances and spoils the targeted optical features~\cite{Garcia2008}.
Although high particle filling fractions (typically $>50\%$) have to be realized, the occurrence of crystalline domains needs to be avoided.
Because of the strong tendency of monodisperse hard spheres to form colloidal crystals, it is very difficult to generate a homogeneous glassy structure.
In addition, optical experiments necessitate macroscopic homogeneity of the material, i.e., crack-free monoliths, because shortcuts (large voids in which the photons propagate ballistically) have to be avoided.
The material should have a minimum optical absorption in the visible region of the spectrum, be it intrinsic or due to impurities.
Because the scattering efficiency increases with the index contrast, compounds with the highest possible refractive index (e.g., titanium dioxide (\tio); $n_\mathrm{rutile} = 2.7$) should be embedded in a low index matrix (e.g., air; $n_\mathrm{air} = 1$).
Accomplishing all these requirements at once is of utmost difficulty and a task for materials science.

PGs macroscopic in size have been rarely reported so far.
Only lower index materials such as polystyrene (PS)~\cite{Garcia2007,Garcia2008,Sapienza2007,Chen2017,Emoto2012}, poly(methyl methacrylate) (PMMA)~\cite{Garcia2007,Sapienza2007}, or silica (SiO$_2$)~\cite{Espinha2016,Montesdeoca2016,RezvaniNaraghi2015}  were studied.
Attempts with higher refractive index materials were done with not perfectly monodisperse zinc sulfide spheres (ZnS, $n_\mathrm{ZnS} = 2.4$) dispersed in deionized water or isopropanol~\cite{Scholz1998}, or with irregularly shaped \tio particles~\cite{Stoerzer2006}.

Resonant light transport behavior in monodisperse PG has been connected to resonant multiple Mie scattering~\cite{Garcia2008,Sapienza2007} but no quantitative description of this connection was given.
In a recent publication, we provided a model able to quantitatively describe light transport in densely packed PGs made of PS spheres~\cite{AubrySchertel2017}.
The presented model of the transport mean free path \ls, the important scattering quantity in multiple light-scattering quantifying inverse turbidity, uses the energy coherent potential approximation (ECPA) for the effective refractive index to account for near field coupling and structural correlations~\cite{Busch1995,*Busch1996}.
In this study~\cite{AubrySchertel2017}, the model was tested against \textit{ab initio} numerical simulations, earlier experimental data obtained from transmission experiments~\cite{Garcia2008}, and experimental results from spectrally resolved coherent backscattering experiments on specially synthesized PS ($n_\mathrm{PS} =  1.6$) colloidal glasses.
In addition, as pointed out recently~\cite{Tiggelen2017}, it is difficult to distinguish between situations where the photon diffusion constant $D$ is small due to Anderson localization effects leading to small \ls , or due to a small transport energy velocity \vE related to Mie resonances.
This long-standing issue of the effect of resonant scattering behavior on dynamic scattering properties---i.e., the photon diffusion constant $D$---and static scattering properties---i.e., the mean-free-path \ls---awaits experimental clarification.

In this paper we first describe the controlled  preparation of crack-free monolithic high-index colloidal \tio PG samples at various refractive indices, particle sizes, and polydispersities. We then determine their turbidities and energy velocities over the entire visible spectrum and compare them to empirically optimized commercial white paints.
Very good overall agreement between measurements and our theoretical model is found.

\section{Fabrication of colloidal photonic glasses}

\begin{figure}
\centering
\includegraphics[]{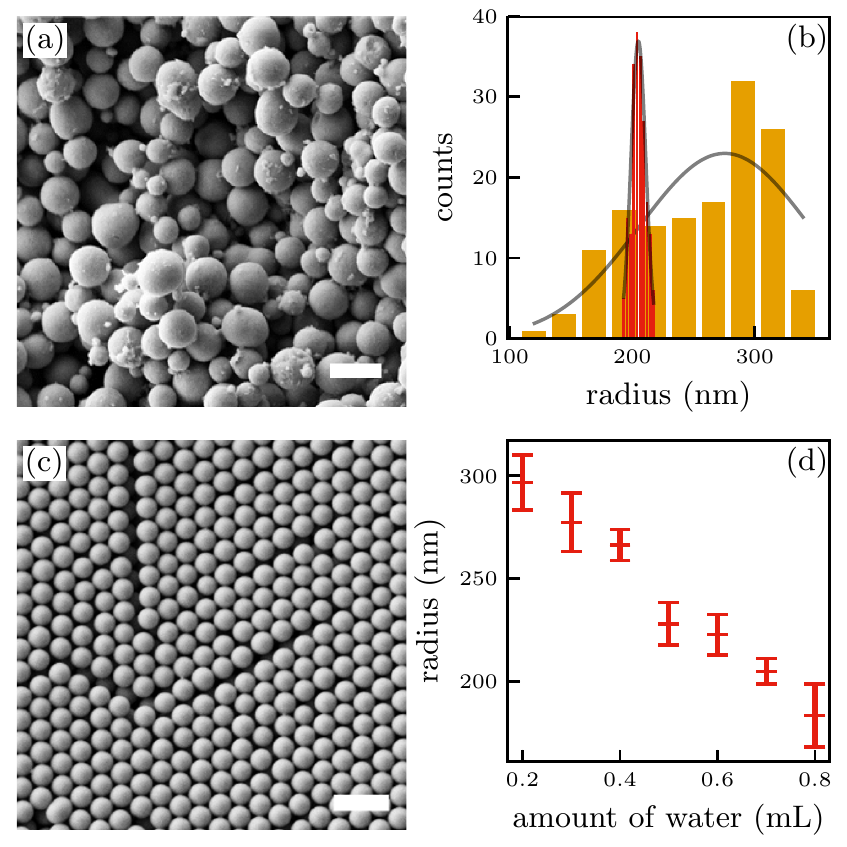}
\caption[]{Scanning electron microscope (SEM) micrographs of \tio colloids.
(a)~polydisperse sample (SD = 22\%),
(c)~monodisperse sample (SD = 2.8\%), scale bars 1~$\mu$m.
(b)~Optimization of polydispersity: Gaussian fits (grey curves) of the particle size distribution functions of the polydisperse sample (orange) SD = 22\% and the monodisperse sample (red) SD = 2.8\%.
(d)~Adjustment of the particle size by addition of deionized water (see Supporting Information~\ref{sec:stericStaParticles}).}
\label{fig:particles}
\end{figure}

The first step is the synthesis of  monodisperse \tio colloids. 
Several reviews have already addressed the preparation of \tio nanoparticles~\cite{Chen2013,Xiang2017}.
Spherical particles made of amorphous titania have been obtained via a sol-gel route~\cite{Barringer1982,Jiang2003}, which is similar to the well-known Stöber method developed for silica~\cite{Stoeber1968}.
Based on previous work from our own group~\cite{Eiden2004}, we have prepared electrostatically stabilized \tio particles first (see Supporting Information~\ref{sec:electrostaticStaParticles}).
Despite numerous attempts it was not possible to prepare samples with sufficiently low polydispersity (see Fig.~\ref{fig:particles}(a,b)).
For instance, the analysis of the sample containing particles with a mean radius $r_\mathrm{colloid} = 240$~nm shows a standard deviation (SD) of~22\% in the size distribution function (grey Gaussian fit to bright histogram in Fig.~\ref{fig:particles}(b)).

The quality of the dispersions could be improved significantly by following a protocol by Tanaka \textit{et al.} on sterically stabilized \tio particles~\cite{Tanaka2009} (see Supporting Information~\ref{sec:stericStaParticles}).
In agreement with these results, we achieved $r_\mathrm{colloid} = 206$~nm particles with a low SD value of only 5.2\%.
We were able to improve the size distribution further by optimization of the reaction parameters while keeping the average size of the particles constant (see \ref{sec:stericStaParticles}).
Finally, particles with $r = 209$~nm and SD = 2.8\% were obtained (Fig.~\ref{fig:particles}(c) and dark histogram in Fig.~\ref{fig:particles}(b)).
As expected, the narrower the size distribution function becomes, the higher is the tendency to form colloidal crystals.
For the optical experiments, it is also important to adjust the mean particle size.
We managed to do so (Fig.~\ref{fig:particles}(d) and Fig.~\ref{fig:histograms}) by the variation of the amount of deionized water used in the synthesis protocol.
The mean particle size was also confirmed by dynamic light scattering.

\begin{figure*}
\begin{center}
\includegraphics[width=0.95\textwidth]{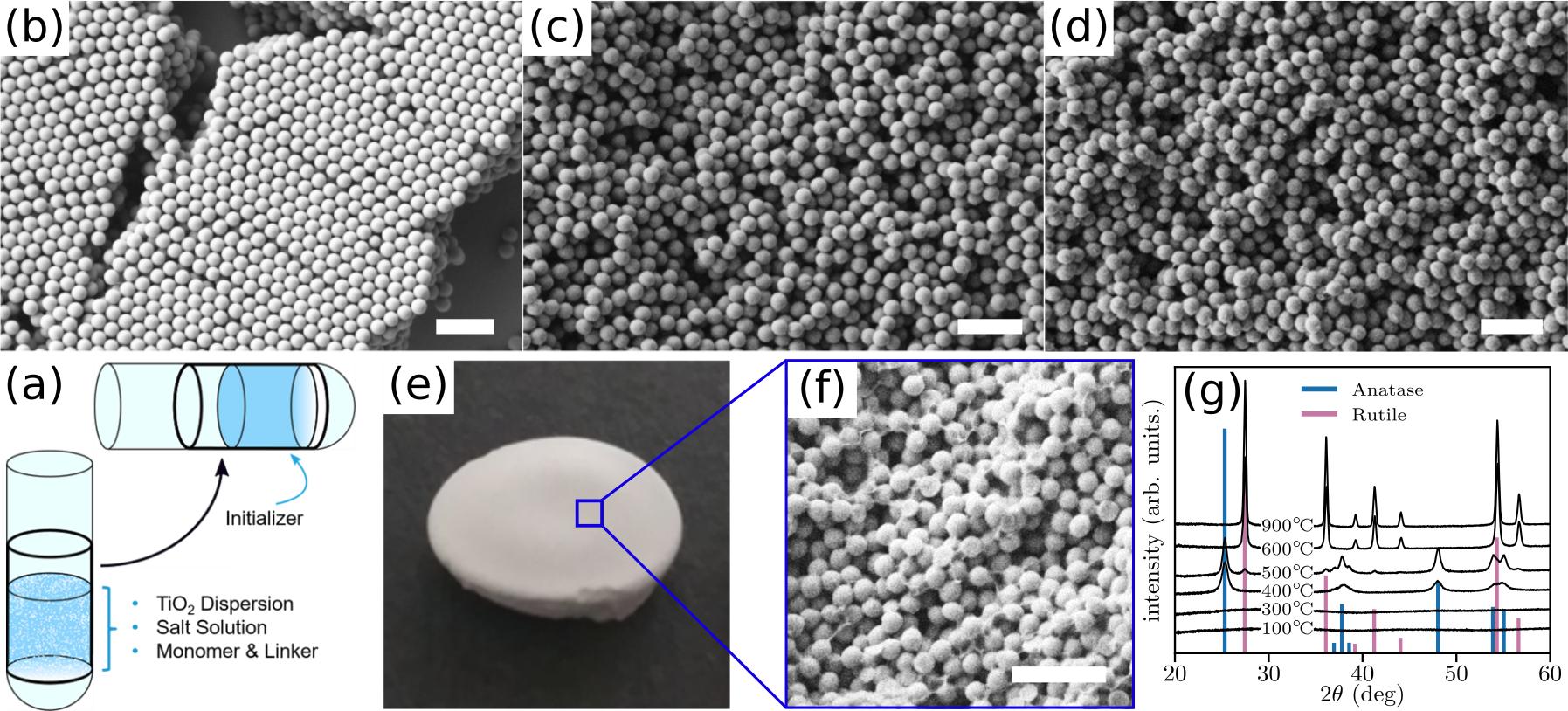}
\end{center}
\caption[]{(a)~Preparation of \tio PG monoliths. SEM micrographs of particles packings with CaCl$_2$ solution added; (b)~no addition, (c)~5\,mM, (d)~10\,mM. (e)~Photographic image and (f)~SEM micrograph of the mesoscopic structure of the final monolith (diameter~$\sim1$\,cm) structurally stabilized by polyacryamide. Scale bars for the micrographs: $2\ \mu$m. (g)~PXRD patterns of materials obtained after sintering at different temperatures. Reference patterns of \tio-anatase (blue bars) and rutile (mauve bars) are also shown.}
\label{fig:samples}
\end{figure*}

Colloidal crystallization at high packing fraction was suppressed during ultracentrifugation of the monodisperse \tio particles dispersions by destabilization with Ca$^{2+}$ (Fig.~\ref{fig:samples}(a)).
This has been shown in previous papers~\cite{Chen2015,*Chen2017a,Chen2017} to be a suitable method for the realization of kinetically disordered packings of PS-latex.
It can be seen that without Ca$^{2+}$, crystallized zones are present (Fig.~\ref{fig:samples}(b)).
The addition of Ca$^{2+}$ has a marked influence and heavily disturbs the emergence of order.
At a concentration of 5\,mM CaCl$_2$, the sample is entirely disordered with no crystalline areas (Fig.~\ref{fig:samples}(c)).
Higher salt concentrations can also be used, but the number of voids increases (Fig.~\ref{fig:samples}(d)), which lowers the filling fraction of the PG.
The resulting monoliths are brittle and difficult to handle.
Removal of the solvent (water) by drying leads to crack formation.
Cracks are undesired because they lead to optical shortcuts.
We therefore stabilized the structure of the monoliths by the \textit{in situ} polymerization of N,N'-methylene-bis-acrylamide.
The resulting polymer is located in between the titania particles and acts like a ``glue'' preserving the structure during drying (see Fig.~\ref{fig:samples}(e,f) and Fig.~\ref{fig:SEMwideview}).
The monoliths are crack-free.
Because the polymer represents only a minor fraction of the entire material and its refractive index ($n_\mathrm{polymer} =  1.4$) is much lower than that of titania, the effect on optical properties may be negligible. The filling fraction was estimated experimentally to be $f\approx0.3\pm0.1$ by tracking the weight percent of particles used and the volume of the sample.
This value for $f$ is close to its optimized value for strong scattering~\cite{Pattelli2018}.

Titania prepared according to sol-gel methodologies is amorphous and should be described as a titanium-oxo-hydroxo phase.
To induce atomic crystallization to anatase or rutile, the materials were sintered at higher temperatures.
According to thermogravimetric analysis (shown in Fig.~S\ref{fig:TGA}) the as-prepared material loses mass in two steps.
In the range $T = 50-200^\circ$C, one sees the removal of solvents and at $T_\mathrm{max} =  401 ^\circ$C, surface bound organics are removed and eventually dehydroxylation takes place~\cite{Wu2017}.
In the FT-IR spectra of a sample treated at $T = 400^\circ$C, one sees neither signal for organic compounds nor for remaining $-$OH groups (see Fig.~S\ref{fig:IR}).
The treatment at higher temperature has the desired effect on the crystallinity of the samples (Fig.~\ref{fig:samples}(g)).
Signals characteristic for \tio-anatase appear in powder x-ray diffraction (PXRD) at a temperature of 400$^\circ$C.
We see the emergence of first peaks indicating rutile phase \tio at 500$^\circ$C. The transformation is finished at 600$^\circ$C, and the entire sample contains only the rutile phase.
The good purity of the materials is also documented by UV-Vis spectra (see~Fig.~S\ref{fig:UVVIS}) and pictures of the anatase and rutile particles are shown in the Supporting Information Fig.~\ref{fig:AnataseRutile}.
The PG monoliths can now be prepared with particles made of amorphous \tio, anatase-\tio and rutile-\tio (see Fig.~\ref{fig:samples}(f)).
This provides a way to tune systematically the refractive index of the photonic material ($n= 2.0, 2.5, 2.7$).

\section{Resonant transport in high index photonic glasses}

Optical turbidity measurements were performed for the samples described above.
For comparison, we measured reference samples consisting of dense packings of commercially available \tio powders (classical white paints) and also compared with our previous data from PS spheres~\cite{AubrySchertel2017}. 
This was done to demonstrate the effect on the scattering behavior of both the refractive index and the monodispersity of the scattering particles.
The scattering strength $\lo/\ls$ is obtained by analyzing the shape of the coherent backscattering cone (CBC)~\cite{Wolf1985,*Albada1985} at different incident wavelengths~\cite{AubrySchertel2017} (\lo is the incident optical wavelength in vacuum).
The width of the CBC is proportional to $1/\kls$ ($k=2\pi/\lo$).
Thus, the width of the CBC increases dramatically for strongly scattering samples such as \tio-based PGs.
Therefore, a large angular range needs to be covered in reflection for determination of \ls of such highly scattering samples.
We used a CBC setup recording backscattering angles up to 60$^\circ$ as described in refs.~\cite{Gross2007} (see Supporting Information~\ref{sec:CBC}).
A tunable laser system (Fianium, WL-SC-400-8 and LLTF VIS) was used as a light source.
For each sample, a wavelength scan from 450\,nm to 780\,nm (limited by the used detectors and optical components) was performed in 10\,nm steps if not mentioned differently.

\begin{figure*}
\begin{center}
\includegraphics[width=0.9\textwidth]{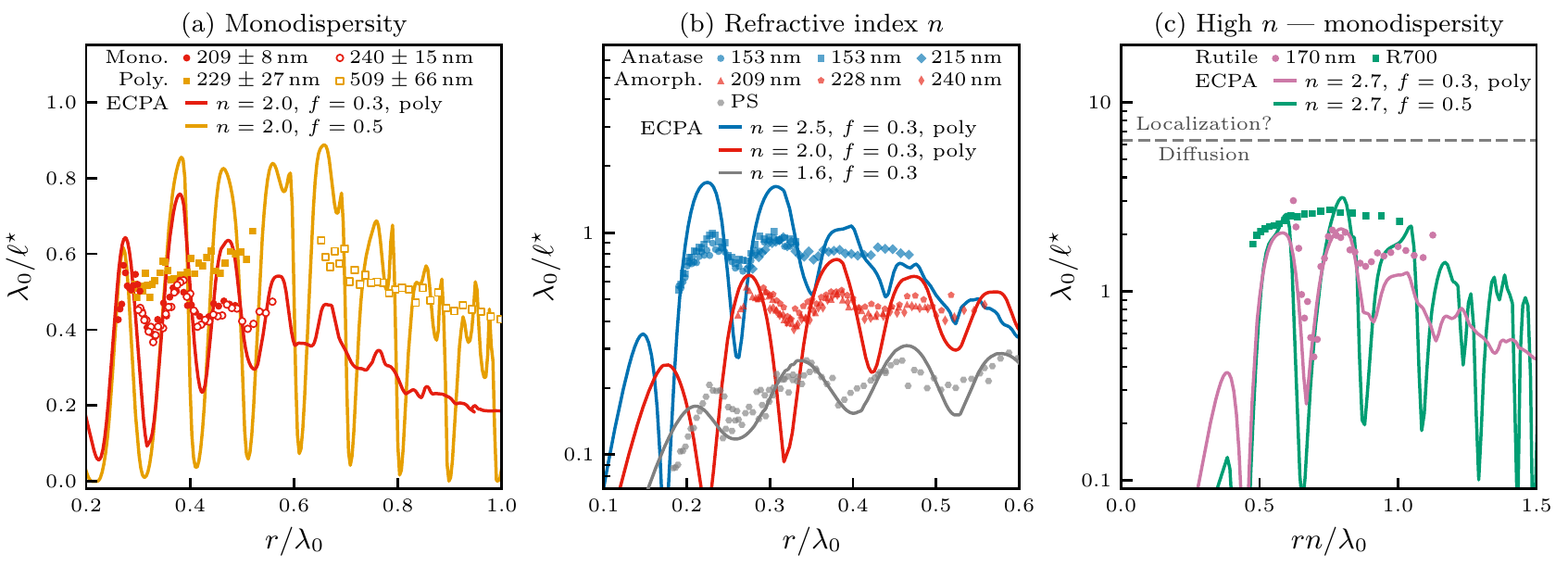}
\end{center}
\caption[]{Comparison of the experimental data (points) with the ECPA scattering model (lines); $f$: filling fraction, $n$: refractive index, poly refers to $5\%$ polydispersity taking into account in the model. (a) PGs prepared from amorphous \tio spheres with different sizes and low (red) and high (orange) polydispersity. (b) Data for PGs prepared from particles with different refractive indices: Polystyrene (grey, data taken from ref.~\cite{AubrySchertel2017}), amorphous-\tio (red), and anatase-\tio (blue). (c) Data for two rutile-\tio based PGs: one prepared from low polydisperse particles ($r = 170\pm5$\,nm, mauve), and one composed of strongly polydisperse commercial \tio (R700, green) as a white paint reference. The dashed grey line indicates the presumed critical value $\kls = 1$ between diffusion and localization.}
\label{fig:ls}
\end{figure*}

In Fig.~\ref{fig:ls}, we plot $\lo/\ls$ for different samples versus the size ratio $r/\lo$ with $r$ the mean particle radius.
The resonant behavior in all measured PGs leads to strong scattering at certain $r/\lo$ values. Note that in the whole paper, each material has its own color (gray: PS spheres, orange/red: amorphous \tio spheres polydisperse/monodisperse, blue: anatase spheres, mauve: rutile spheres, green: R700).
The experimental data are compared with a model developed by us recently~\cite{AubrySchertel2017}, which is based on the calculation of the scattering cross section, taking into account the Mie solution~\cite{Bohren1998} and the glass structure factor~\cite{Percus1958,Fraden1990} (see Supporting Information~\ref{sec:ECPA}).
Because of the close proximity of the individual scatterers,  considering the usual far field Mie solution of a single scatterer with refractive index $n$ in a background with refractive index \no does not properly describe the multiple scattering behavior.
At high packing fractions, due to the electromagnetic coupling and to the positional correlations of neighboring particles, the single scattering Mie resonances are shifted.
One way to approximate this \textit{a priori} complex scattering is to renormalize \no to an effective refractive index \neff, which is calculated using the energy-density coherent-potential approximation (ECPA) introduced by Soukoulis \textit{et al.}~\cite{Busch1995,*Busch1996}
The key step to describe the effective medium is to introduce a coated sphere as a basic scattering unit. 
The particle shell has a refractive index \no and a thickness related to the average particle distance in the glass and couples electromagnetically each scatterer with the surrounding effective medium.
The ECPA-based scattering model was already tested successfully by us on PGs made of PS spheres~\cite{AubrySchertel2017}.
Unlike other models used so far, it predicts very well the resonant behavior in low index PGs without any adjustable parameter.
We now check the predictions of this model on \tio-based high-index PGs.
Figure~\ref{fig:ls}(a) illustrates the influence of the monodispersity of the particles on the scattering behavior of the PGs.
All samples contain amorphous \tio spheres but they differ regarding particle size and polydispersity.
The orange data on the figure shows the spectral measurements of the scattering strength of two PGs prepared from \tio colloids (polydispersity $< 5\%$) with $r = 209\pm8$ and $240\pm15$\,nm, respectively.
These data are compared to the ECPA model for $ n_\mathrm{TiO_2} = 2.0 $, a filling fraction $f = 0.3$ and a polydispersity of 5\%.
There is good agreement between experimental data and our model.
The positions of the experimentally observed resonances overlap very well with the predictions from the ECPA scattering model.
The amplitudes of the observed resonances are weaker than expected from the model, but the mean absolute values match quite well.
Note here that in the experiments the amplitude might be somewhat lowered because of the spectral width of the light source (2.5\,nm).
Another reason is the polymer network that holds the particles together which may reduce slightly the index contrast.
The data as well as the model show that for larger $r/\lo$ values, the resonances smear out.
This is explained by the residual polydispersity~\cite{AubrySchertel2017}. 
The latter effect is more striking when using a less monodisperse titania dispersions ($r = 229\pm27$, $509\pm66$\,nm) to synthesize polydisperse PGs.
Because ordered structures do not form for such high polydispersity (Fig.~\ref{fig:particles}(b)) these PGs were prepared by simple compression of the synthesized and dried powders (see Supporting Information~\ref{sec:PGR700}).
The red data in Fig.~\ref{fig:ls}(a) show again a direct comparison between experimental data and the corresponding ECPA model.
As expected, in this case resonances are absent over the entire $r/\lo$ range; however, the data follow the trend of the model.

In the following, we focus on monodisperse samples and analyze the effect of the refractive index (Fig.~\ref{fig:ls}(b)).
The resonances of the material based on amorphous \tio (red) are enhanced at least by a factor two compared to particles with lower refractive index (PS, grey) (note the logarithmic plot of the scale).
The refractive index is higher for anatase-\tio (blue) spheres, and the resulting PGs show an even more enhanced scattering and more defined resonances.
Again, there is a very good agreement between the experimental data and the ECPA model.

Rutile-\tio particles have the highest refractive index. Because commercial white paints contain irregular-shaped rutile-\tio particles, it is worth using these as a reference system~\cite{Stoerzer2006,Aegerter2006,Sperling2013}.
PGs were obtained by compacting commercially available \tio particles to tablets with $f\approx0.5$ (see~\ref{sec:PGR700}).
The powders have high polydispersities in the range $25-47\%$ with a mean size varying from 170 to 283\,nm (DuPont R700; see~\ref{sec:PGR700}).
The data are plotted in Fig.~\ref{fig:ls}(c) versus the relative size parameter $r n_\mathrm{TiO_2} /\lo$.
This index correction is performed as these data were obtained by varying the wavelength while the scattering model is calculated for a fixed wavelength (590\,nm) varying the particle radius.
As the refractive index is wavelength dependent, the used wavelength in the experiments are weighted by the wavelength dependent $n$ taken from ref.~\cite{DEVORE1951}.
Note that this correction is small and was neglected for PS, amorphous, and anatase \tio due to a lack of literature data.
The scattering strength of the commercial rutile powders follows the trend of the ECPA scattering model over a large range of $r n_\mathrm{TiO_2} /\lo$ from 0.5 to 1.0 (green points in Fig.~\ref{fig:ls}(c)).
Due to the arguments given above, resonances are absent.
Note that the amplitude of the scattering strength is higher than expected by the model mean value.
Either the model slightly underestimates the scattering strength or the random shape of the particles (different from a PG made out of polydisperse spheres) leads to stronger scattering.

We now discuss the scattering features of the optimum sample (monodisperse rutile-\tio - based PGs).
The mauve points in Fig.~\ref{fig:ls}(c) show the results for $r = 170\pm5$\,nm.
The scattering behavior of the monodisperse PG follows remarkably well the predictions from the model without any fit parameter.
Mean free path values of $\ls \sim 230-240$\,nm are reached corresponding to $\lo/\ls\sim3.13$.
This is comparable to the values reported for the strongest scattering commercial white paints~\cite{Stoerzer2006} where no indications of Anderson localization were found~\cite{Sperling2016}.
Note that these values are still larger than the critical  value $\kls \approx 1$  were the transition to Anderson localization is expected from the Ioffe-Regel criterion (Fig.~\ref{fig:ls}(c); dashed line)~\cite{IoffeRegel}.
Therefore, an important outcome of our investigation is that the regime of light localization cannot be reached even with the optimized spherical scattering systems presented here.

\section{Energy transport velocity}

Yet another unsolved issue in the quest for Anderson localization is the respective role of the static time averaged scattering strength $1/\ls$ and the dynamic diffusion constant, which depends---in addition to $\ls$---on the energy velocity \vE  of the light wave through $D = \vE \ls/3$ in a multiple scattering sample. \vE can thus be calculated from $D$, which is accessible via photon time of flight (ToF) measurements~\cite{Drake1989,Sperling2016}.
How \vE is affected by resonant scattering, which of these two quantities (\vE and \ls) dominates $D$ and what controls the Anderson transition is not fully understood in the literature~\cite{Busch1995, Tiggelen2017}.
We therefore performed measurements of ToF combined with \ls to characterize the resonant transport behavior in a PG formed by amorphous \tio spheres with $r = 228$\,nm.
ToF is done by sending a short laser pulse onto the sample and by measuring the transmitted intensity using a time-resolved photodetector.
For a diffusive sample, the diffusion constant  $D$ is extracted by fitting such ToF curves with diffusion theory, provided the sample thickness $L$ is known~\cite{Drake1989,Sperling2016}.

\begin{figure}
\begin{center}
\includegraphics[width=0.9\columnwidth]{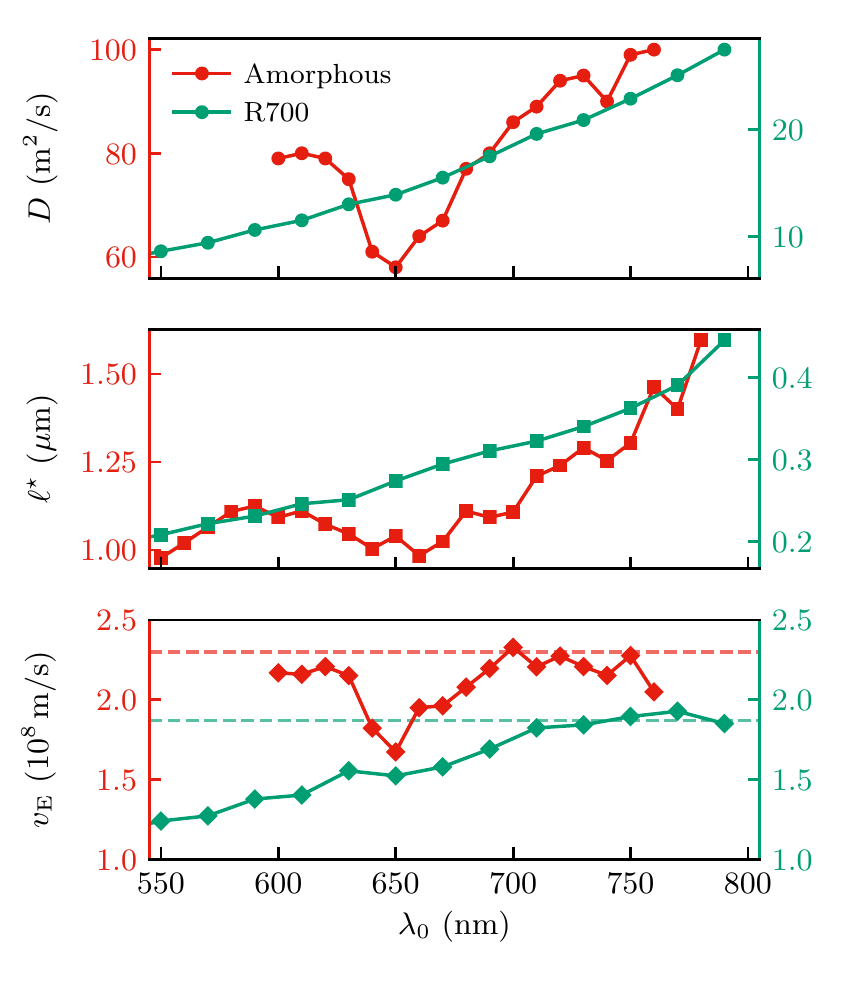}
\end{center}
\caption[]{Spectral measurements of (a) the diffusion constant $D$, (b) the transport mean free path \ls, and (c) the calculated energy velocity \vE, for (red, left scales) an amorphous \tio PG with particle radius $r = 228$\,nm and a sample thickness $L =0.5$\,mm, and (green, right scales) for a commercial rutile \tio powder having a sample thickness $L=0.4$\,mm (DuPont R700).
The dashed lines in (c) indicate the energy velocity calculated by the Maxwell-Garnett effective refractive index (note here the common left and right scales).}
\label{fig:comparisonvE}
\end{figure}

Figure~\ref{fig:comparisonvE}(a) shows measurements of the diffusion constant $D$ versus the wavelength $\lo$.
For the amorphous PG (red points), $D$ has a minimum at exactly the same position where \ls has its minimal value (Fig.~\ref{fig:comparisonvE}(b)).
The calculated values of \vE (Fig.~\ref{fig:comparisonvE}(c)) can be compared to $\vE = c/n_\mathrm{MG} \simeq 2.3\times 10^8$\,ms$^{-1}$ (dashed red line), with $c$ the velocity of light and $n_\mathrm{MG}$ the Maxwell-Garnett effective refractive index~\cite{Garnett1904}.
Accounting for the error in the measurement of the sample thickness ($L = 0.5\pm0.2$\,mm), the observed energy velocity agrees quite well with that value.
Though the data are somewhat noisy, a minimum in \vE can be identified close to the minimum in \ls.
Therefore, the behavior observed for \ls, $D$ and \vE are strongly correlated.

To reveal the origin of the resonances, we also measured commercial \tio powders to study the dynamic transport behavior in samples made out of randomly shaped, polydisperse particles for comparison.
The green points in Fig.~\ref{fig:comparisonvE} (R700, $L=0.40\pm0.05$\,mm) show one example
illustrating that the diffusion constant $D$ decreases with decreasing incident wavelength.
As expected from the ill-defined shape of the particles, no resonant behavior is observed.
The energy velocity \vE decreases monotonically with increasing size ratio in contrast to the behavior of \ls.
Moreover, \vE is lower than the value expected from $\vE=c/n_\mathrm{MG} \sim 1.87\times 10^8$\,m/s, using $f=0.5$ in the calculation of $n_\mathrm{MG}$ (green dashed line in Fig.~\ref{fig:comparisonvE}(c)).
All this shows the complexity of the scattering behavior in the case of multiple scattering from materials composed of randomly shaped, densely packed particles, which is clearly beyond the applicability of the model presented here.

\section{Conclusion}

Understanding the optical properties of white paints and appreciating the potential of PGs relies on materials suitable as model systems for quantitative comparison of a manageable theoretical model with experiments.
In this paper we describe the preparation of highly monodisperse titania colloids with adjustable refractive index by controlling the particle's crystallinity (amorphous vs  anatase vs rutile).
A centrifugation method combined with colloidal destabilization was successful in suppressing any colloidal crystallization and large monolithic PG samples were obtained.
These allowed for accurate measurements of the scattering strength and of the diffusion constants over a wide range of particle size parameters covering the whole visible spectrum.
We studied the influence of colloidal particle size, polydispersity, and refractive index in comparison to our recently proposed model for resonant light transport in densely packed sphere systems~\cite{AubrySchertel2017}.
Good agreement of data and model was found, both for the amplitudes and spectral resonance positions of turbidity and energy velocity. This paper therefore paves the way toward a class of controlled and application-optimized photonic materials where isotropy is elemental. 

Even for close-to-perfect PGs made from monodisperse spherical rutile-\tio, light transport is dictated by diffusion: the conditions for reaching strong localization of light are not met.
To reach this goal, other disordered photonic structures than just randomly packed spheres have to be invented.
One recently proposed way is to use hollow or coated spheres~\cite{Escalante2017} to hinder the propagation of longitudinal evanescent fields responsible for new transport channels~\cite{Skipetrov2014PRL,RezvaniNaraghi2015}.
Another possible direction may be suggested by nature where, despite the low refractive index material in the exoskeleton of some white beetles (chitin, $n=1.6$), light scattering is optimized in an unprecedented way~\cite{Vukusic2007,*Burresi2014}. The Cyphocilus beetle uses elongated scatterers to achieve highly scattering material, similar to recently published work on artificial fibrillar networks~\cite{Toivonen2018, *Syurik2018}.
One could imagine reaching Anderson localization of light with a photonic structure similar to what is found in the white beetle, but made of high-index materials such as rutile-\tio.

\begin{acknowledgments}

The samples synthesis was developed and carried out by I.W., while the light-scattering experiments were done by L.S.

We thank the Deutsche Forschungsgemeinschaft for funding within Project No. Ae94/2-4, Ma817/8 and within the framework of the Collaborative Research Centre SFB-1214, Project No. A5 and Particle Analysis Center (PAC). The Schweizer Nationalfonds Grant No. SNF 200020M-162846 is acknowledged as well as the Center for Applied Photonics (CAP) and the Zukunftskolleg (Independent Research Starting Grant) of the Universität Konstanz.

\end{acknowledgments}

\bibliography{Bibliographie}

\onecolumngrid

\appendix

\pagebreak
\setcounter{figure}{0}
\renewcommand{\thefigure}{S\arabic{figure}}
\renewcommand{\theHfigure}{Supplement.\thefigure}

\onecolumngrid
        
\vspace*{2em}
\begin{center}
    \large \bf Supporting Information
\end{center}

\vspace{2em}
The Supporting Information contains information on the TiO$_2$ colloid preparation and characterization, how photonic glasses are prepared from TiO$_2$ colloids and commercial TiO$_2$ powders, details on the coherent backscattering measurements and on the computation of the scattering strength.

\vspace{5em}

\twocolumngrid



\section{TiO2 colloid preparation}
\subsection{Electrostatic stabilized particles}
\label{sec:electrostaticStaParticles}
For the preparation of the titanium dioxide (\tio) particles, 20\,mL ethanol and $80-100\,\mu$L of a 0.1\,M KCl solution were mixed together and stirred for several minutes.
0.6\,mL of the Ti(O\,$^\mathrm{i}$Pr)$_4$ precursor was added to the solution with 1.5\,mL per minute.
After 4 hours, the reaction was stopped and the particles were collected by filtration, washed with ethanol, and dried under ambient conditions at 60$^\circ$C.

\subsection{Steric stabilized particles}
\label{sec:stericStaParticles}
\begin{figure*}
\subfigure[Synthesis with 0.2 mL H$_2$O: $\left<r\right> = 296.8$\,nm. Scale bar in the SEM micrograph: 1 $\mu$m.]{
\centering
\includegraphics[width=0.85\textwidth]{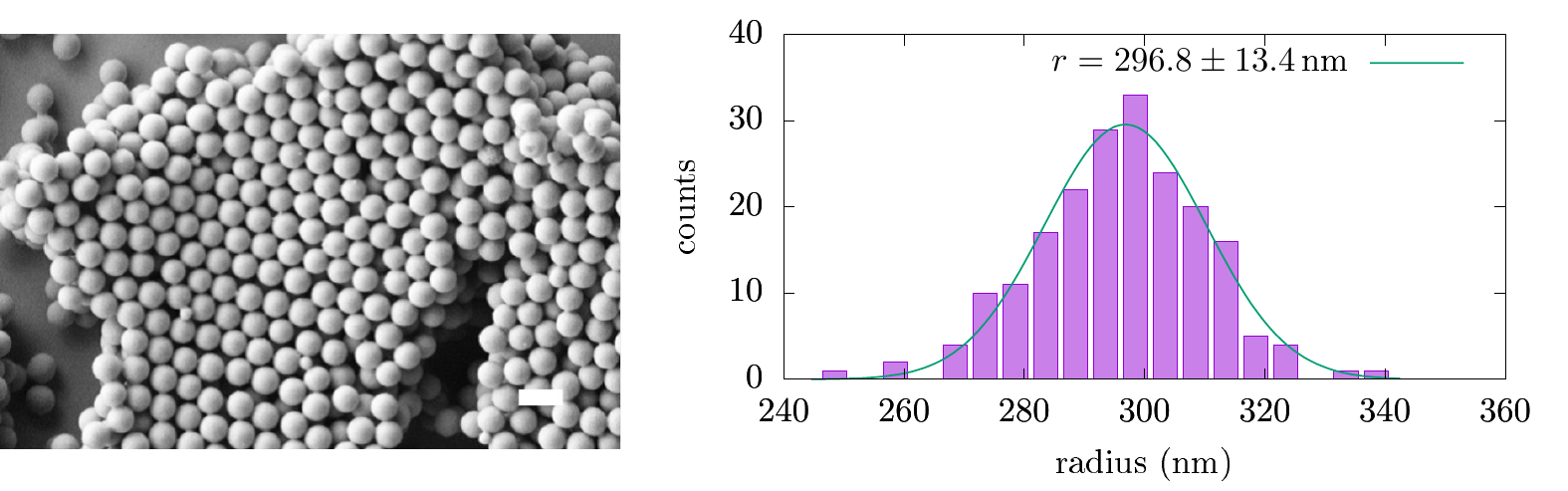}
\label{fig:0p2}
}
\subfigure[Synthesis with 0.5 mL H$_2$O: $\left<r\right> = 228.0$\,nm. Scale bar in the SEM micrograph: 2 $\mu$m.]{
\centering
\includegraphics[width=0.85\textwidth]{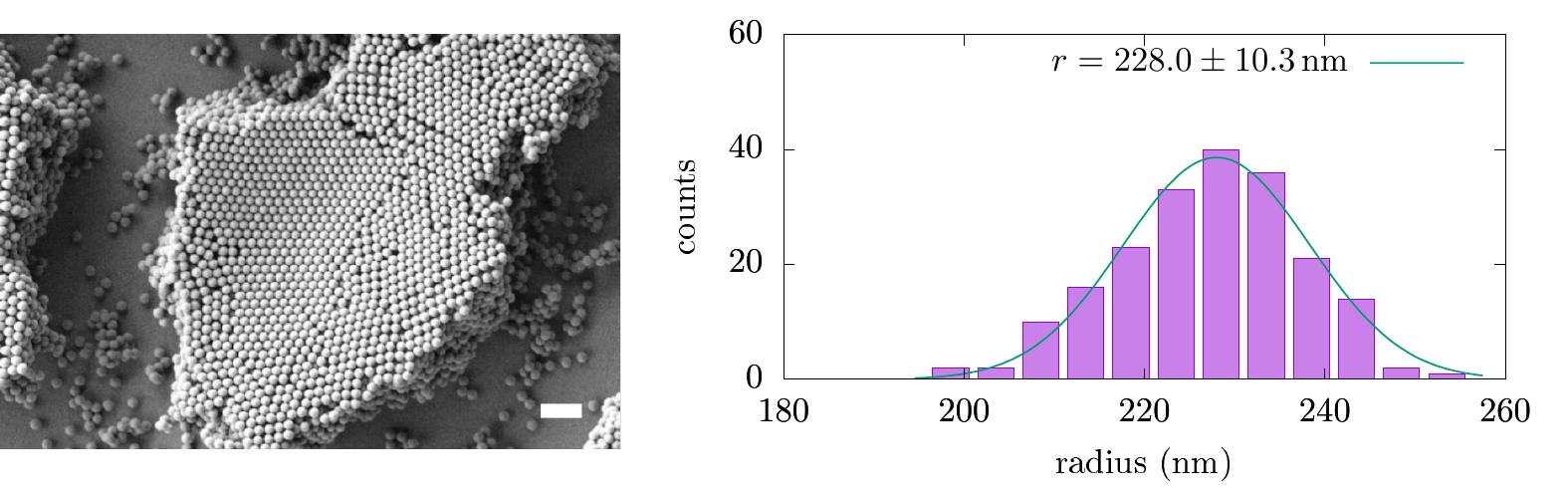}
\label{fig:0p5}
}
\subfigure[Synthesis with 0.7 mL H$_2$O: $\left<r\right> = 204.9$\,nm. Scale bar in the SEM micrograph: 1 $\mu$m.]{
\centering
\includegraphics[width=0.85\textwidth]{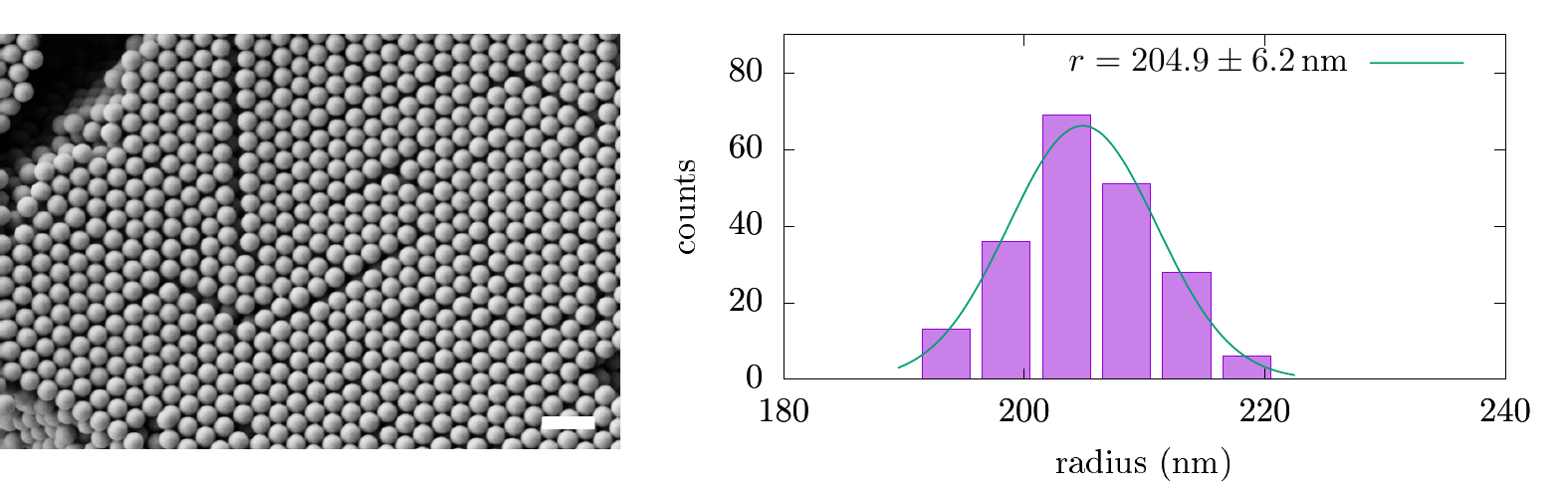}
\label{fig:0p7}
}
\caption{Variation of mean particle size of \tio particles for different amounts of H$_2$O.
Left: SEM micrographs; right: histograms of 200 counted particles.
Bin size in the histograms: 5\,nm. The fits are Gaussian fits.}
\label{fig:histograms}
\end{figure*}

\begin{figure*}
\subfigure[TGA]{
\centering
\includegraphics[height=5cm]{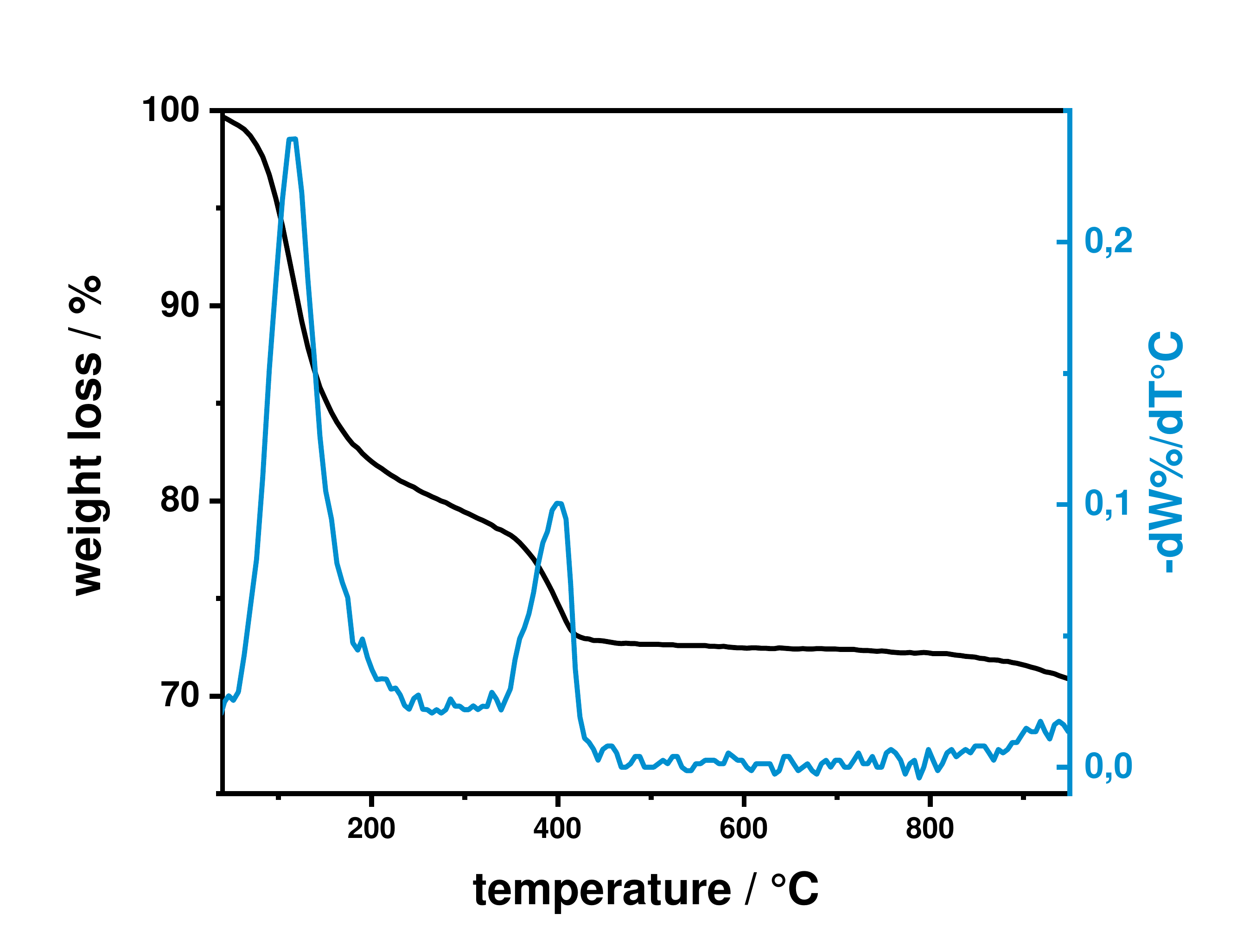}
\label{fig:TGA}
}
\subfigure[FT-IR]{
\centering
\includegraphics[height=5cm]{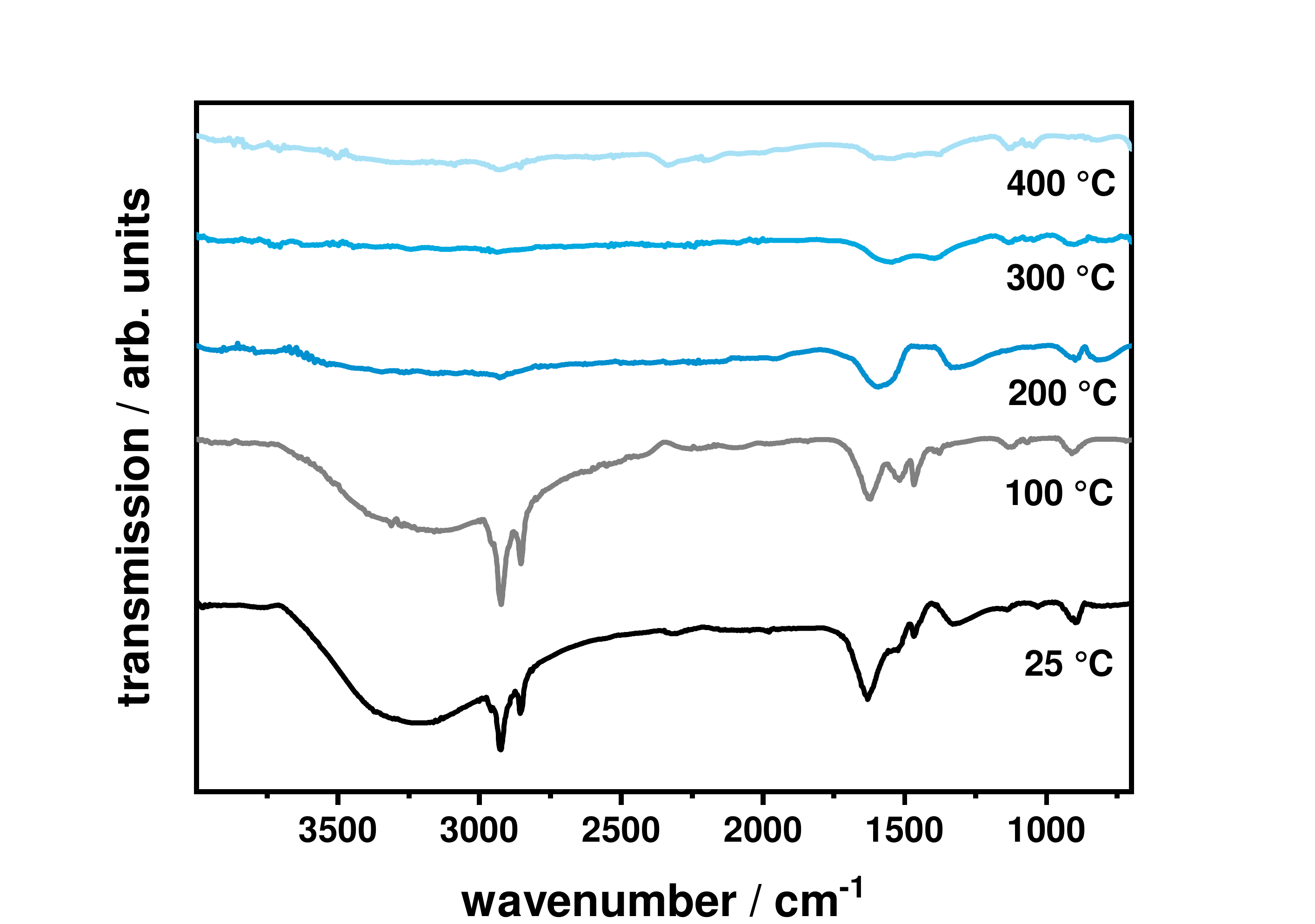}
\label{fig:IR}
}
\subfigure[UV-Vis]{
\centering
\includegraphics[width=0.9\columnwidth]{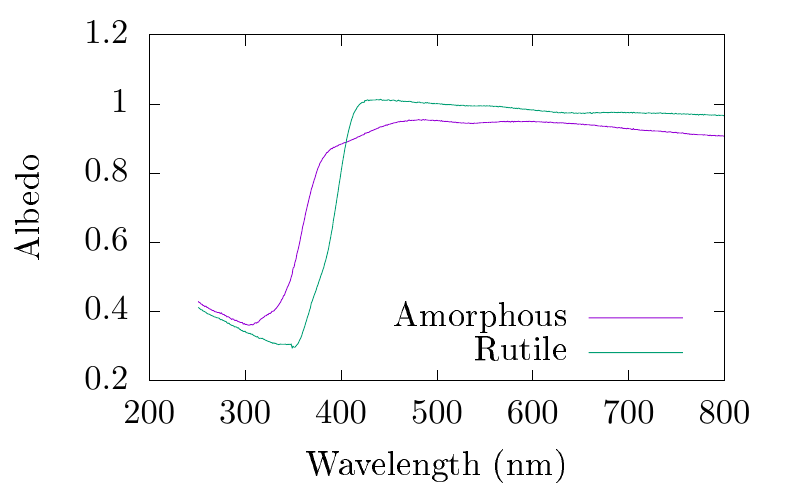}
\label{fig:UVVIS}
}
\caption{(a) TGA trace (black) and first derivative (blue).
(b) FT-IR spectra of samples treated at different temperatures.
(c) UV-Vis diffuse reflectance acquired in an integrating sphere containing either amorphous or rutile \tio.}
\label{fig:TGAIR}
\end{figure*}

In a typical synthesis, 1\,g of dodecylamine was dissolved in a mixture of 106\,mL of methanol and 42\,mL of acetonitrile.
Afterwards, 0.2 to 0.7\,mL deionized water was added, and the solution was stirred for 10 minutes.
1\,mL of the Ti(O\,$^\mathrm{i}$Pr)$_4$ precursor was added at once to the solution.
After stirring for 8 hours, the obtained particles were washed 3 times with methanol and for further analysis dried at 60$^\circ$C under ambient conditions.
The size of the colloids were determined by measuring the radius of 200 particles on Scanning electron microscope (SEM) pictures (Fig.~\ref{fig:histograms}).
The thermogravimetric analysis (TGA), Fourier-transform infrared (FT-IR) spectra of a sample treated at different temperatures and UV-Vis spectra are plotted in Fig.~\ref{fig:TGAIR}. Prior to the heat treatment, different features can be observed in the FT-IR spectra. The broad signal with its maximum at $\nu = 3200\,\mathrm{cm}^{-1}$ corresponds to the oxygen-hydrogen vibration of the remaining ethanol and the nitrogen-hydrogen vibration of the dodecylamine. The two sharp bands at $\nu = 2857\,\textrm{cm}^{-1}$ and $\nu = 2925\,\textrm{cm}^{-1}$ are due to the asymmetric and symmetric carbon-hydrogen vibration. For increasing temperatures these bands are decreasing and at 200$^\circ$C  they completely disappear. At $\nu = 1620\,\textrm{cm}^{-1}$ the Ti-O-H stretching vibration is observable, but at 400$^\circ$C, it has completely vanished.

\begin{figure*}
\centering
\includegraphics[width=0.8\textwidth]{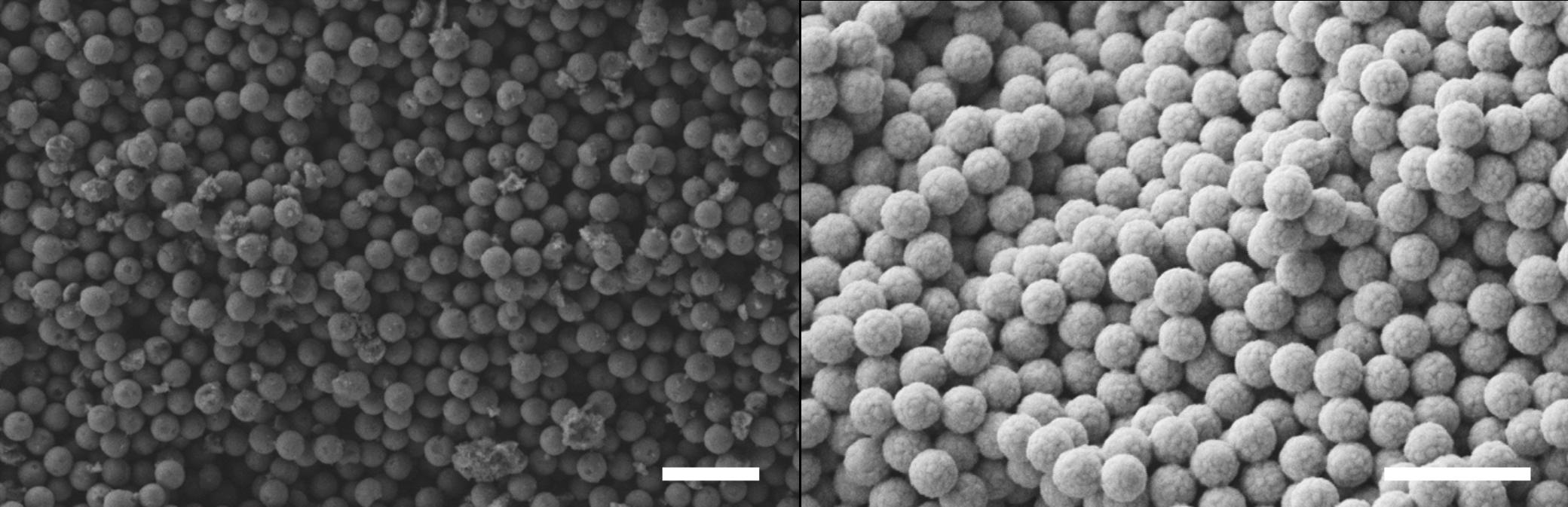}
\caption{(left) Anatase and (right) rutile particles after sintering. (SEM micrographs, scale bar 1\,$\mu$m)}
\label{fig:AnataseRutile}
\end{figure*}

Anatase particles were obtained through sintering at 400$^\circ$C for 4 hours while sintering at 700$^\circ$C for 1 hour led to rutile particles (see main text Fig.~2(g)).
Fig.~\ref{fig:AnataseRutile} shows SEM picture of such anatase and rutile particles.

\section{Preparation of photonic glasses }
\subsection{Photonic glasses (PGs) of \tio colloids}
\begin{figure*}
\centering
\includegraphics[width=0.8\textwidth]{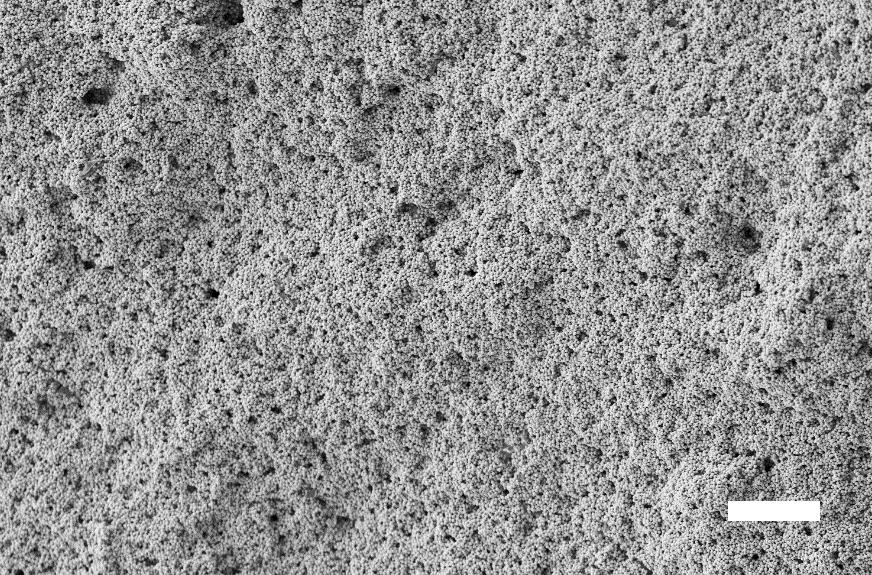}
\caption{Polymer-stabilized PG monolith of titania nanospheres (SEM micrographs, scale bar 10\,$\mu$m)}
\label{fig:SEMwideview}
\end{figure*}

For preparation of a free-standing monolith, 1\,mL of the \tio dispersion, 0.36\,mL of a 30\,wt\% acrylamide solution, 0.32\,mL of a 2\,wt\% N,N'-methylenbisacrylamide solution, 0.04\,mL H$_2$O and 0.08\,mL of a 0.18\,M CaCl$_2$ solution were mixed in a glass vial with flat bottom.
The solution was centrifuged for 1 hour, 1000 revolutions per minute (rpm) and 25$^\circ$C.
Afterwards the supernatant solution was removed and 0.02\,mL N,N,N',N-tetramethylethan-1,2diamine and 0.02\,mL of a 10\,wt\% ammonium persulfate solution were added to initiate the hydrogel's polymerization.
The monoliths were dried at 80$^\circ$C for 1 hour to obtain a free-standing monolith. 
Figure~\ref{fig:SEMwideview} shows a wide view of a typical sample.

\subsection{PGs of polydisperse \tio powders}
\label{sec:PGR700}
\begin{figure*}
\centering
\includegraphics[width=0.8\textwidth]{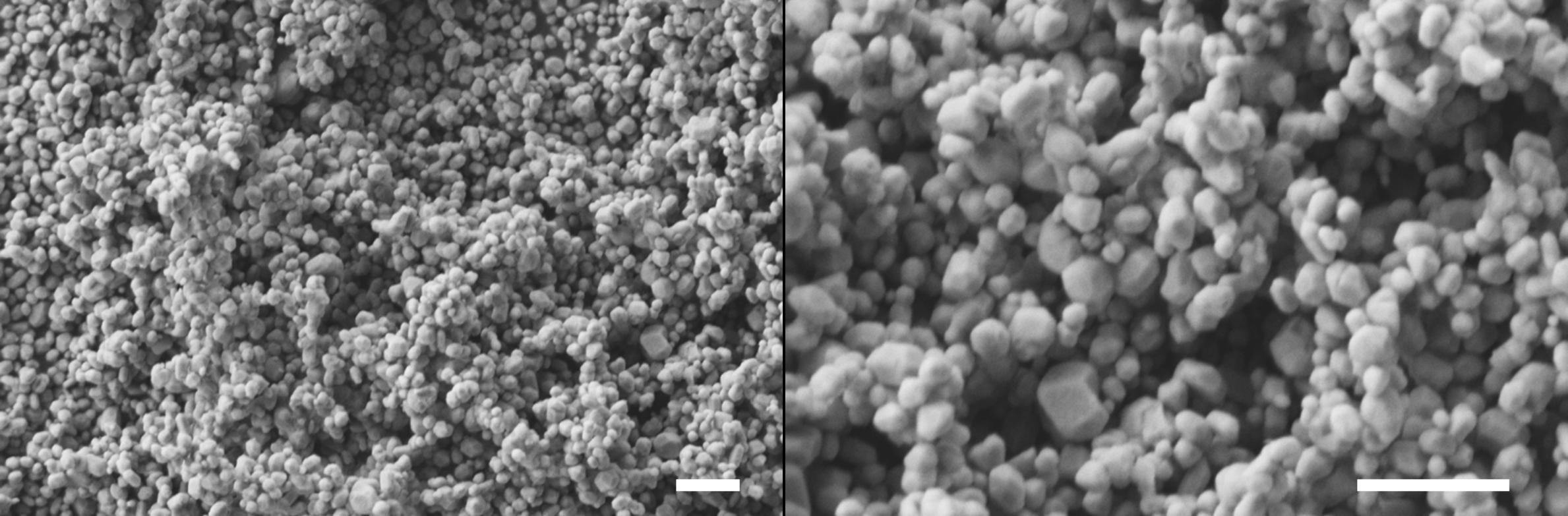}
\caption{Reference materials prepared by compressing commercial DuPont R700 \tio powders.
(SEM micrographs, scale bars 1\,$\mu$m)}
\label{fig:R700}
\end{figure*}

Fig.~\ref{fig:R700} shows the shape of the polydisperse R-700 commercial \tio colloids.
To obtain densely packed powders in slab geometry, the polydisperse powders are pressed to tablets in cylindrical shaped sample holders.
A lid can be screwed on the cylindrical holder before the powder is filled in the holder and pressed with a stamp fixed by screws.
This leads to filling fractions of approximately $f=0.5$.
The powders are pressed between two glass plates to perform transmission experiments.
One glass plate can be removed from the holder for coherent backscattering cone (CBC) experiments to guarantee for a flat sample/air surface to avoid additional reflections by the glass.

\section{Materials structure characterization}
\begin{itemize}
\item SEM measurements were performed using a Zeiss CrossBeam 1540XB.
\item Powder X-Ray Diffraction (PXRD) measurements were performed using a Bruker D8 DISCOVER with Cu K$_\alpha$ radiation.
\item TGA and Differential scanning calorimetry measurements were performed using a Netzsch STA F3 Jupiter.
\item FT-IR spectra were measured with a PerkinElmer Spectrum 100 spectrometer with an attenuated total reflection (ATR) unit.
\item Transmission electron microscope (TEM) measurements were performed using a Jeol JEM 2200 FS.
\item UV-Vis measurements were performed using Varian Cary 100 Scan UV-Vis Spectrometer with Ulbricht-sphere.
\item Preparative Ultracentrifugation was done using an Optima XL-I 70 of Beckman Coulter. The rotor was a SW 55 Ti rotor.
\end{itemize}

\clearpage
\section{Coherent backscattering cone measurements}
\label{sec:CBC}
We used a coherent backscattering cone setup recording angles from $-60^{\circ}$ to $60^{\circ}$.
A white light laser (Fianium, WL-SC-400-8 ) with a total output power of $P\approx 8$~W in a wavelength range of $400-2500$\,nm was used as a light source. The white light pulses (repetition rate of 80~MHz) are coupled in a spectral tunable filter (Fianium, LLTF VIS) that uses two spectral overlapping Bragg gratings to tune the wavelength between $400-1000$\,nm with a spectral width of 2.5\,nm and an average output of $4\,\mathrm{mW}/\mathrm{nm}$. The beam is focused through an entrance hole in the half circle mount (1.2\,m in diameter) of the 256 photodiodes which are the detectors~\cite{Gross2007}. The beam is focused with the focal point at the diodes position to minimize the non-measurable angles in the backscattering direction. At small angles ($<9.75^{\circ}$) photodiode arrays (Hamamatsu, S5668) are used for a higher resolution of $0.15^{\circ}$ at the cone tip while at larger angles ($<19.55^{\circ}$) single photodiodes (Hamamatsu, S4011) are used with a resolution of $0.7^{\circ}$. For even larger angles ($<60^{\circ}$) the same diodes are positioned with a resolution of $\approx 1^{\circ}$. The illuminating beam passes a circular polarizer before impinging onto the sample. The filter wavelength range ($380-780$\,nm) limits the spectral measurements. Another circular polarizer foil in front of the photo diodes filters single scattering events. The sample is mounted on a rotating motor to average out the static speckle pattern during the measurement. The illumination spot is 1\,cm in diameter. A teflon sample is used for calibration as it has such a narrow cone that only the incoherent background is detected by the diodes.

\section{Computation of the scattering strength}
\label{sec:ECPA}
As explained in our previous paper~\cite{AubrySchertel2017}, the transport mean free path \ls is computed from
\begin{equation}
\ls= \frac{\ell_\mathrm{s}}{1-\langle \cos \theta \rangle} \overset{\text{sphere}}{=}  \frac{1}{1-\langle \cos \theta \rangle  }\frac{4 \pi r^3}{3f \sigma_{\mathrm{s}}}. \label{eq:lstar}
\end{equation}
where $\ell_\mathrm{s}$ is the photon scattering mean free path, $r$ is the radius of the colloids forming the photonic glasses, $f$ is the filling fraction of the colloids in the glass, and $\sigma_{\mathrm{s}}$ and $\langle \cos \theta \rangle$ are computed from the standard formula~\cite{Fraden1990}
\begin{equation}
\sigma_\mathrm{s} = \frac{\pi}{k^2}\int_0^\pi F(\theta) S(\theta) \sin \theta \,\dd\theta,
\label{eq:scattCS}
\end{equation}
\begin{equation}
\langle \cos \theta \rangle= \dfrac{\int_0^\pi \cos \theta F(\theta) S(\theta) \sin \theta \,\dd\theta}{ \int_0^\pi F(\theta) S(\theta) \sin \theta \,\dd\theta },
\label{eq:asyPara}
\end{equation}
where $k=2\pi/\lambda$ with $\lambda$ the wavelength of the light in the surrounding medium, $F(\theta)$ is the Mie form factor~\cite{Bohren1998} and $S(q)$ is the Percus-Yevick structure factor~\cite{Percus1958} of the glass (with $q=2k\sin\theta/2$).
In these formula, the near field effects are taken into account by replacing all wave vectors $k$ by $k_{\mathrm{eff}}=2\pi n_{\mathrm{eff}}/\lambda_0$ with $n_\mathrm{eff}$ the energy coherent potential approximation (ECPA) refractive index~\cite{Busch1995,*Busch1996}.
The ECPA effective refractive index depends on $r$, $f$ and on the refractive indices of the particles $n_\mathrm{p}$ and of the surrounding matrix (air, $n_0$), and is computed iteratively as described in our previous paper~\cite{AubrySchertel2017}.

In the end, the so-computed transport mean free path $\ls$ depends on $\lambda_0$ (wavelength in vacuum), $r$, $f$, $n_\mathrm{p}$ and $n_0$.
As already shown in our previous work~\cite{AubrySchertel2017}, the positions of the resonances do not depend on $f$.

\onecolumngrid
%
%
%

\end{document}